\documentclass[12pt]{article}
\usepackage{graphicx}
\usepackage{amsmath}

\newcommand\pubnumber{DPF2015-262}
\newcommand\pubdate{\today}

\def\umn{
$^{1}$Department of Physics, Panjab University, Chandigarh - 160014, India\\
$^{2}$Department of Physics and Astronomy, University of South Carolina, Columbia, SC 29208, USA}
\def\Title#1{\begin{center} {\Large #1 } \end{center}}
\def\Author#1{\begin{center}{ \sc #1} \end{center}}
\def\Address#1{\begin{center}{ \it #1} \end{center}}

\newcommand\pubblock{\rightline{\begin{tabular}{l} \pubnumber\\
         \pubdate  \end{tabular}}}
\newenvironment{Abstract}{\begin{quotation}  }{\end{quotation}}
\newenvironment{Presented}{\begin{quotation} \begin{center} 
             PRESENTED AT\end{center}\bigskip 
      \begin{center}\begin{large}}{\end{large}\end{center} \end{quotation}}

\def\beq{\begin{equation}}
\def\eeq#1{\label{#1}\end{equation}}
\def\eeqn{\end{equation}}

\def\beqa{\begin{eqnarray}}
\def\eeqa#1{\label{#1}\end{eqnarray}}
\def\eeqan{\end{eqnarray}}

\let\bar=\overbar

\def\Dslash{\not{\hbox{\kern-4pt $D$}}}
\def\dslash{\not{\hbox{\kern-2pt $\del$}}}

\def\msb{{\bar{\ssstyle M \kern -1pt S}}}

\begin{document}
\begin{titlepage}
\pubblock

\vfill
\Title{Neutrino Flux Studies at NO$\nu$A}
\vfill
\Author{Kuldeep Kaur Maan$^{1}$,  Hongyue Duyang$^{2}$, Sanjib Ratan Mishra$^{2}$ $\&$ Vipin Bhatnagar$^{1}$}
\Address{\umn}
\center{(for the NO$\nu$A Collaboration)}

\vfill
\begin{Abstract}
We present the systematic-error study of the neutrino flux in the NO$\nu$A experiment.   
Systematic errors on the flux at the near detector (ND), far detector (FD), and 
the ratio FD/ND, due to the beam-transport and hadro-production 
are estimated.  Prospects of constraining the $\nu_{\mu}$ and $\nu_{e}$ 
flux using data from ND are outlined. 
\end{Abstract}
\vfill
\begin{Presented}
DPF 2015\\
The Meeting of the American Physical Society\\
Division of Particles and Fields\\
Ann Arbor, Michigan, August 4--8, 2015\\
\end{Presented}
\vfill
\end{titlepage}
\def\thefootnote{\fnsymbol{footnote}}
\setcounter{footnote}{0}

\section{Introduction}

The NuMI Off-axis $\nu_{e}$ Appearance (NO$\nu$A) experiment is composed of two functionally identical 
detectors.  NO$\nu$A is designed to address a broad range of open questions in the neutrino sector through precision measurements of $\nu_{\mu} \rightarrow \nu_{e}$, $\bar{\nu}_{\mu}\rightarrow \bar{\nu}_{e}$, $\nu_{\mu} \rightarrow \nu_{\mu}$ and $\bar{\nu}_{\mu} \rightarrow \bar{\nu}_{\mu}$ oscillations including neutrino mass hierarchy, CP violation in the neutrino sector. 
To minimize the systematic errors, a functionally identical Near Detector (ND ~ 0.3kt) is placed close to the neutrino source, while the far detector (FD ~ 14kt) is located 810km from the source,
 observes the oscillated beam.
For all of the oscillation measurements, NOvA takes
advantage of a two-detector configuration to mitigate
uncertainties in neutrino flux, neutrino cross sections,
and event selection efficiencies.  NO$\nu$A uses Fermilab's NuMI beam line as its neutrino source. 
This paper focuses on the estimating the systematic errors on the NOvA flux, and constraining 
these errors using the ND-measurements.

\section{\bf  Detectors}
\begin{figure}[ht]
\begin{center}
\includegraphics[width=14cm]{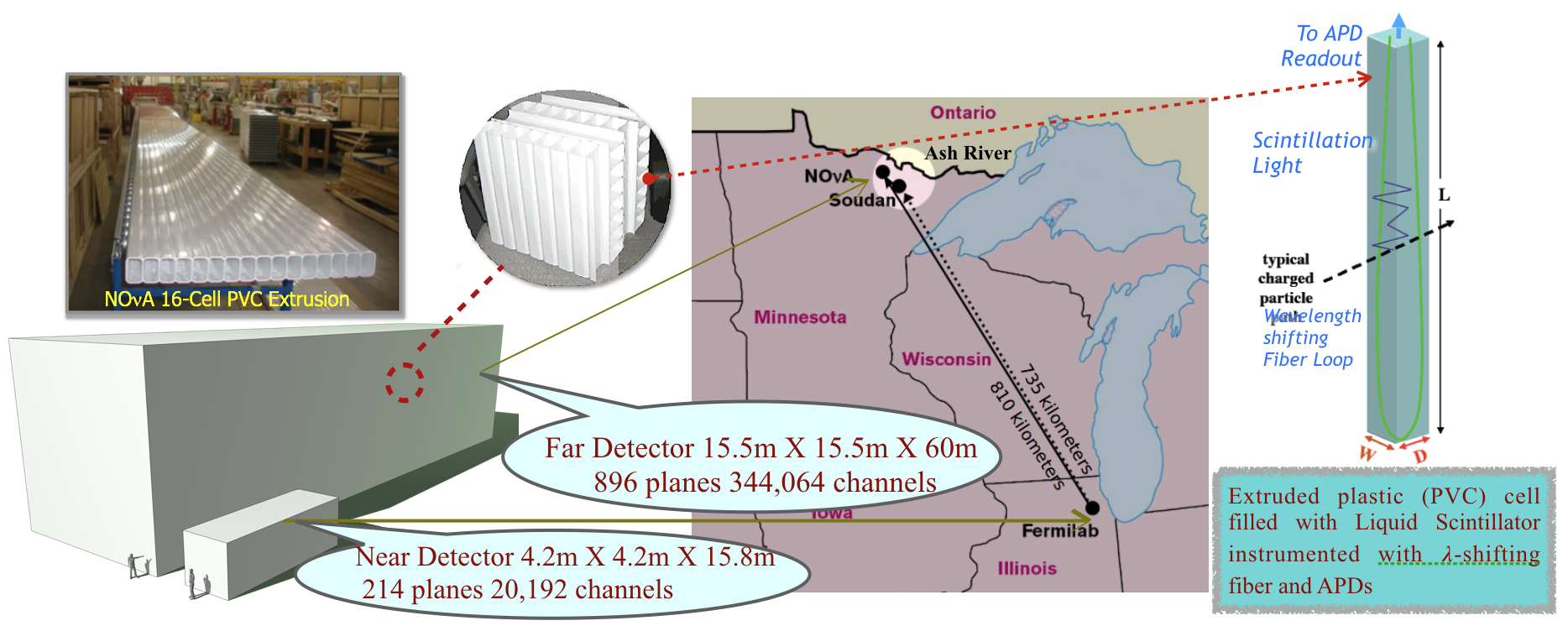}
\caption{NOvA detectors, with a human figure shown for scale. The FD differs from the ND only in the length of its PVC cells and the number of layers present.}\label{f1}
\end{center}
\end{figure}
The NO$\nu$A detectors are situated 14 mrad off the NuMI beam axis, so they are exposed to a relatively narrow band of neutrino energies centered at ~2 GeV. The NO$\nu$A detectors are 
largely active (~65\%) and  highly segmented detectors composed of low-Z tracking calorimeters. The segmentation and the overall mechanical structure of the detectors are provided by a lattice of PVC cells, as shown in Figure~\ref{f1}.  The dimension of the PVC cells is 4$\times$6 cm$^2$. 
Each layer is 0.15 X0 (radiation- length) thick. Each plane is composed of individual cells instrumented with 1-sided readout using avalanche photodiodes (APD).  Each layer in the detectors is oriented orthogonally to adjacent ones to provide 3D event reconstruction, a cut-away view of the PVC cellular structure. The ND(FD), where the cells are 4.2~m (15.5~m) long, 
is composed of 192 (896) planes. 

\section{\bf  NuMI Beam Line}
\begin{figure}[ht]
\begin{center}
\includegraphics[width=14cm]{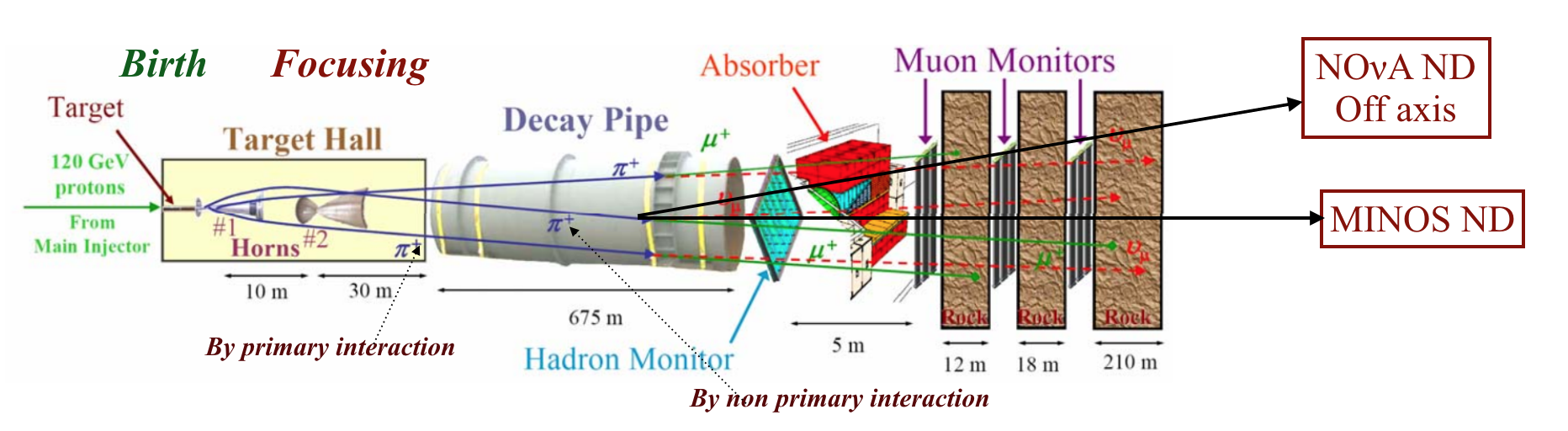}
\caption{Schematic of the NuMI Beam: Shown are the primary proton-C collision, the $\pi^{+}$, $\pi^{-}$, K$^{+}$, K$^{-}$, and K$_{L}^{0}$ mesons that are the primary progenitor of neutrinos, the focusing beam elements, and secondary/tertiary sources of neutrinos. }\label{fig:p2}
\end{center}
\end{figure}

\begin{figure}[ht]
\begin{minipage}{.4\linewidth}
  \includegraphics[width=\linewidth]{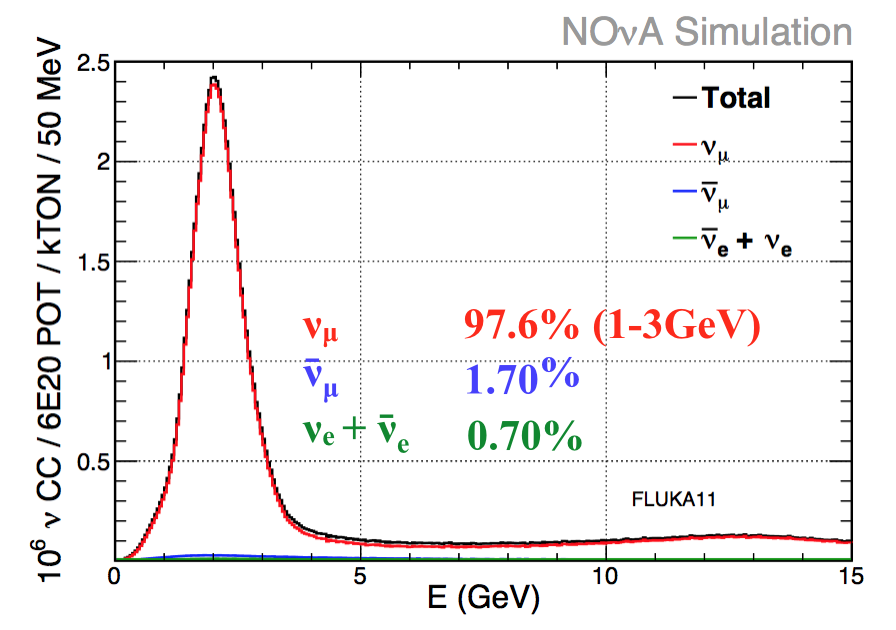}
  \caption{$\nu$ mode: horns focus positives}\label{fig:p3}
 
\end{minipage}
\hspace{.08\linewidth}
\begin{minipage}{.4\linewidth}
  \includegraphics[width=\linewidth]{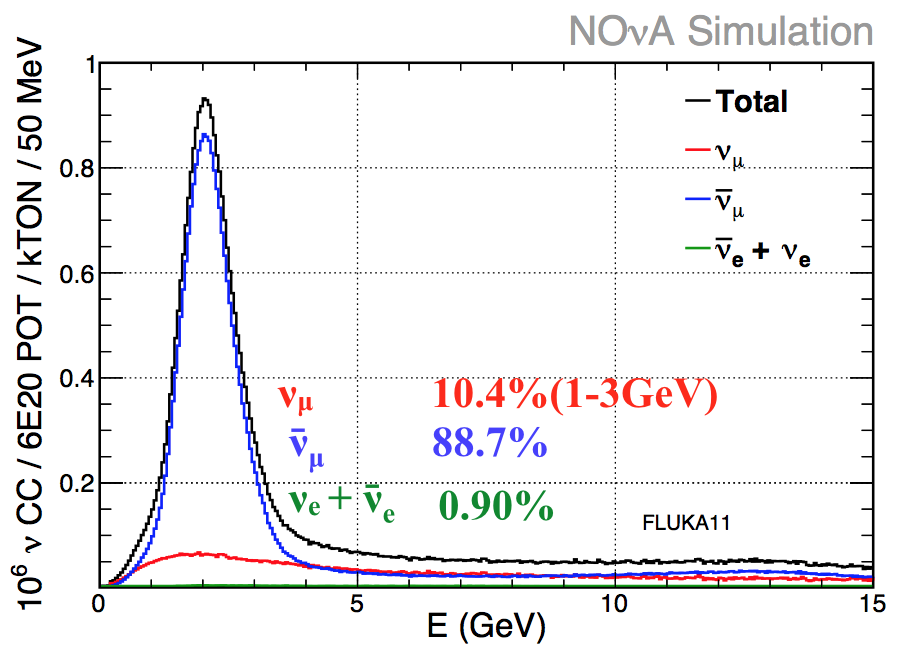}
  \caption{$\bar{\nu}$ mode: horns focus negatives}\label{fig:p4}
 
\end{minipage}
\end{figure}

The schematic of the NuMI beamline is shown in Figure~\ref{fig:p2}. 
A 120 GeV proton-beam  from the Main Injector is impinged upon a graphite target. 
Secondary particles produced from the p-C interaction are focused by two horns where a strong magnetic field is present. Of all these secondary particles, most important are pions and kaons, because they are the dominant source of neutrinos. After being focused, they are left free to decay in a decay pipe. At the end of the decay pipe, the  hadrons  are absorbed in a hadron-absorber\cite{beam}. Due to Off-axis position of detector,  the beam is rich in pure $\nu_{\mu}$ in neutrino mode as shown in Figure~\ref{fig:p3} and $\bar\nu_{\mu}$ in the antineutrino mode, see Figure~\ref{fig:p4}.

\section{\bf  Motivation for the Beam-Systematics}
\begin{figure}[ht]
\begin{center}
\includegraphics[width=10cm]{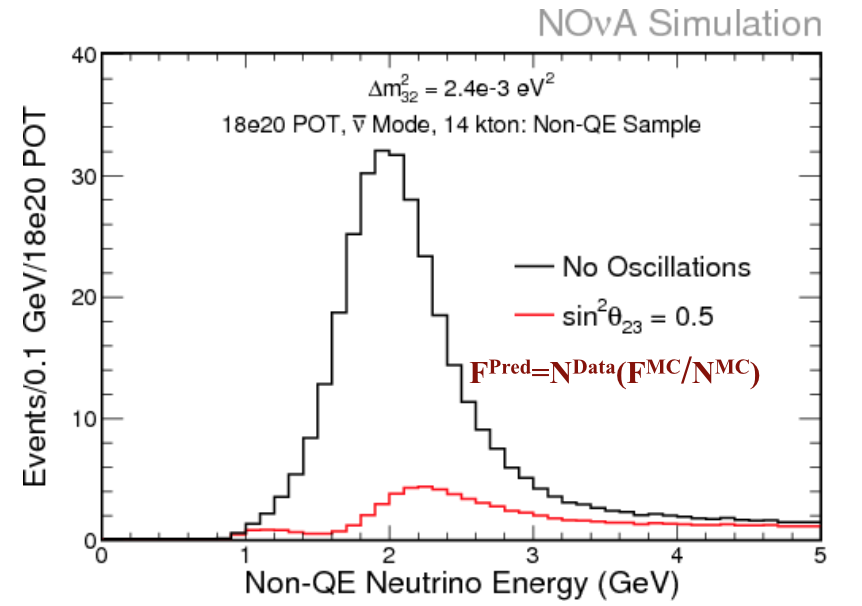}
\caption{Oscillated and Un$\mbox{-}$Oscillated spectrum at FD}\label{fig:p5}
\end{center}
\end{figure}
The sensitivity of the oscillation studies critically depends upon the precise prediction of the 
ratio of the unoscillated to oscillated flux, $\nu_{\mu}$, $\bar{\nu_{\mu}}$, $\nu_{e}+\bar{\nu_{e}}$ in  FD/ND(E$_{\nu}$) as shown in Figure~\ref{fig:p5}. Uncertainties in FD/ND come from the proton-nucleon hadro-production and the beam transport simulation. Needed are data-driven methods to constrain the uncertainties. The most important data are the NO$\nu$A-ND data. Other constraints include MINOS, NDOS (Near Detector Prototype On Surface) data, and the hadro-production data (MIPP, NA49...)\cite{NumiSyst}.

\section{\bf Beam systematic uncertainties}

Neutrino flux prediction based solely upon MC is not precise. Large uncertainties
associated with the proton-nucleon hadro-production processes in primary and
secondary/tertiary targets induce a large uncertainty ($\approx$20-25\%)  in the neutrino flux. 
However measurements and discoveries of the elements of the neutrino
mixing matrix critically depend upon the precision with which one can predict the
neutrino flux ratio at the far detector (FD) with respect to the near detector (ND) as a
function of the neutrino energy ($E_{\nu}$) and the $\nu_{e}$/$\nu_{\mu}$ flux ratio. Poor measurements of the secondary meson production in p-Nucleus collision, 
d$^{2}\sigma$/dx$_{F}$dp$_{T}$ 
contribute to the flux error. The mesons include 
 $\pi^{\pm}$, K$^{\pm}$, and K$^{0}$, produced in the 120 GeV p-C collision. 
  Additional errors are due to the beam-transport simulation. 
  Beam Simulation is based upon 
  FLUGG 2009.4 Flugg 2009-3d\cite{Flugg} and Fluka (2011.2b.6) as standard Monte Carlo.  
  

\subsection{Beam Transport Systematics}
\label{subsec:2}

 This study includes variations in  parameters associated with the beam transport 
 and presenting the variations in the flux  at ND, FD, and (FD/ND) as a function 
 of neutrino energy\cite{lisa}\cite{tom}.  
For the beam simulation, the nominal parameters are: Flugg 2009-3d and Fluka (2011.2b.6),  Forward Horn Current, nominal Horn Current 200kA, linear BField distribution. Beam spot size 1.1mm, PEANUT generator turned on for all energies\cite{fullstat}.
 \begin{figure}[ht]
\begin{center}
\includegraphics[width=14cm]{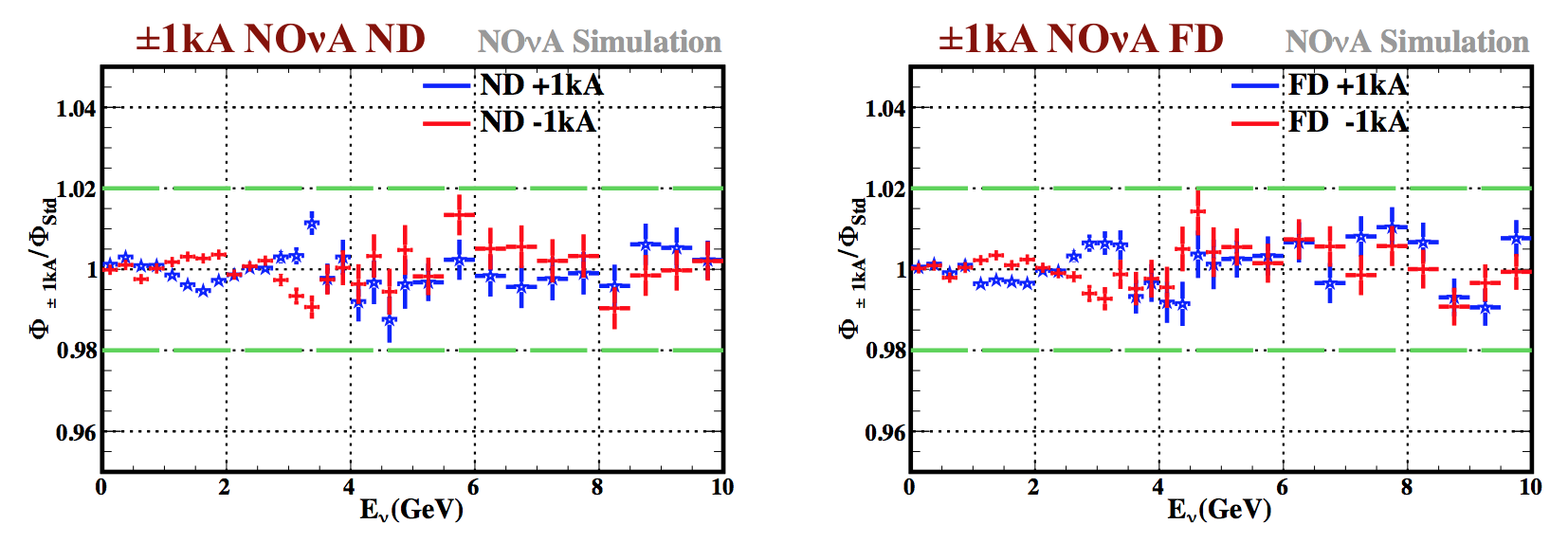}

\caption{Left: ratio of $\nu$ flux with variants, $\pm$1kA shift, blue (+1kA) and red (-1kA), to nominal $\nu$ flux (200kA) at NO$\nu$A ND. Right: ratio of $\nu$ flux with variants,$\pm$1kA shift, blue (+1kA) and red (-1kA), to nominal $\nu$ flux (200kA) at NO$\nu$A FD. ).}\label{fig:p6}

\end{center}
\end{figure}
 \begin{figure}[ht]
\begin{center}
\includegraphics[width=14cm]{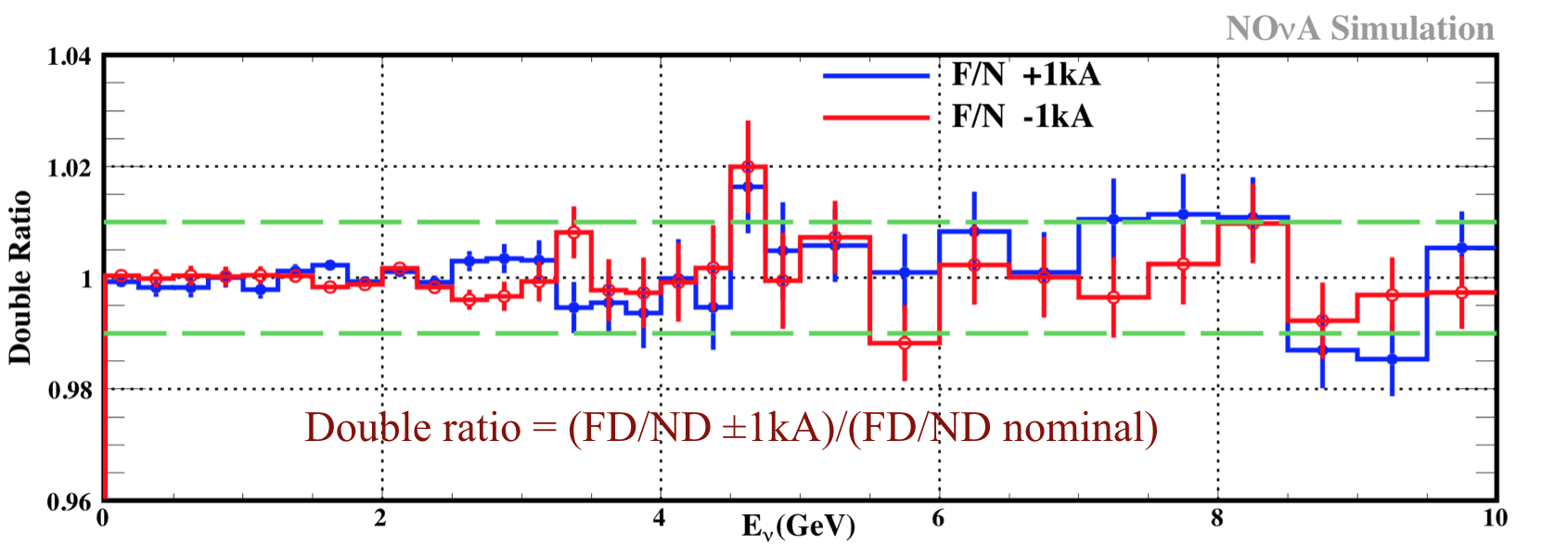}
\caption{Double-Ratio $\frac{\Phi_{FD/ND}(variant)}{\Phi_{FD/ND}(Std)}$ for +1kA shift (blue) and -1kA shift (red)}\label{fig:p7}

\end{center}
\end{figure}
\begin{table}[!tbp]
\caption{$\delta(\%)$ $\nu_{\mu}$ and Energy(Mean and  RMS) at NO$\nu$A ND (1-3)GeV}
\label{tab:4}       
%
%
\begin{tabular}{p{5.5cm}p{3cm}p{2cm}}
\hline\noalign{\smallskip}
Shift &$\delta(\%)$ at ND & $\delta(\%)$ at FD \\
Std   & 0.00 & 0.00   \\
+1kA Horn Current &-0.20&-0.16\\
-1kA  Horn Current &0.16&0.10\\
Horn1 +2mm X $\&$ Y&-0.44&-0.39\\
Horn1 -2mm X $\&$Y&-1.70&-1.76\\
Horn2 +2mm X $\&$ &-0.51&-0.47\\
Horn2 -2mm X $\&$ Y&0.37	&0.30\\
Exp Magnetic Field &-4.30&-4.32\\
BmPosX  0.5mm &-0.66	&-0.68\\
BmPosX  -0.5mm &0.26&0.24\\
BmPosY  0.5mm&0.13&0.18   \\
BmPosY -0.5mm&-0.35&-0.45 \\
BmSpotSize +0.2mm X $\&$ Y&-0.77&-0.81 \\
BmSpotSize -0.2mm X $\&$ Y &0.29&0.29\\
TarPos +2mm in Z &-0.08&-0.09\\
FTFP$\_$BERT &-3.65&-3.76\\
\noalign{\smallskip}\hline\noalign{\smallskip}
\end{tabular}

\end{table}

\noindent
{\bf Flux Systematics Variants for Beam transport:} The following variations are considered: \\
$\bullet$ Horn Current shifted by  $\pm$kA w.r.t nominal\\ 
$\bullet$ Beam spot size shifted by $\pm$.2mm both in X and Y w.r.t nominal\\ 
$\bullet$ Horn1 $\&$ Horn2 position shifted by $\pm$2mm w.r.t nominal\\ 
$\bullet$ Target position shifted by +2mm shift w.r.t nominal\\
$\bullet$ Beam position on the target shifted by $\pm.5$mm in X $\&$ Y separately\\ 
$\bullet$ B-field modeling changed to an exponential magnetic field  (0.77cm skin depth) in the horn skin.

 In the following we show only one sample flux-error calculation due to the variation in 
 the horn current,  shown in Figure~\ref{fig:p6} and Figure~\ref{fig:p7}.  Similar set of
calculation is performed for each of the errors listed above. The fractional variations in the 
yield of $\nu_\mu$ at ND and FD due to the various error-conditions are presented 
in table~\ref{tab:4}\cite{technote}.

\subsection{Uncertainties in hadron production based on NA49 data }
\label{subsec:3}
One key systematic uncertainty is the uncertainty on the simulation of the production of pions and kaons off the carbon target because the yield and kinematics of the 
pions and kaons coming off the target can alter the abundance and energy of
focused pions and kaons that decay to produce muon and electron neutrinos
observed at the NO$\nu$A detectors.
 The core concept is to vary hadro-production parameters within reasonable limits in a
 physically justifiable way, to use the shifted hadron production to create  shifted-flux distributions, 
 and then use the shifted-flux
distributions to generate a covariance matrix. 
Alternative hadron production parameterizations are created
 around a best fit (BMPT\cite{bmpt}) to a FLUKA simulation of the NA49 target; the resulting 
 error covers the difference between the Fluka-MC and NA49\cite{na49}. 
 
\begin{figure}[ht]
  \includegraphics[width=14cm]{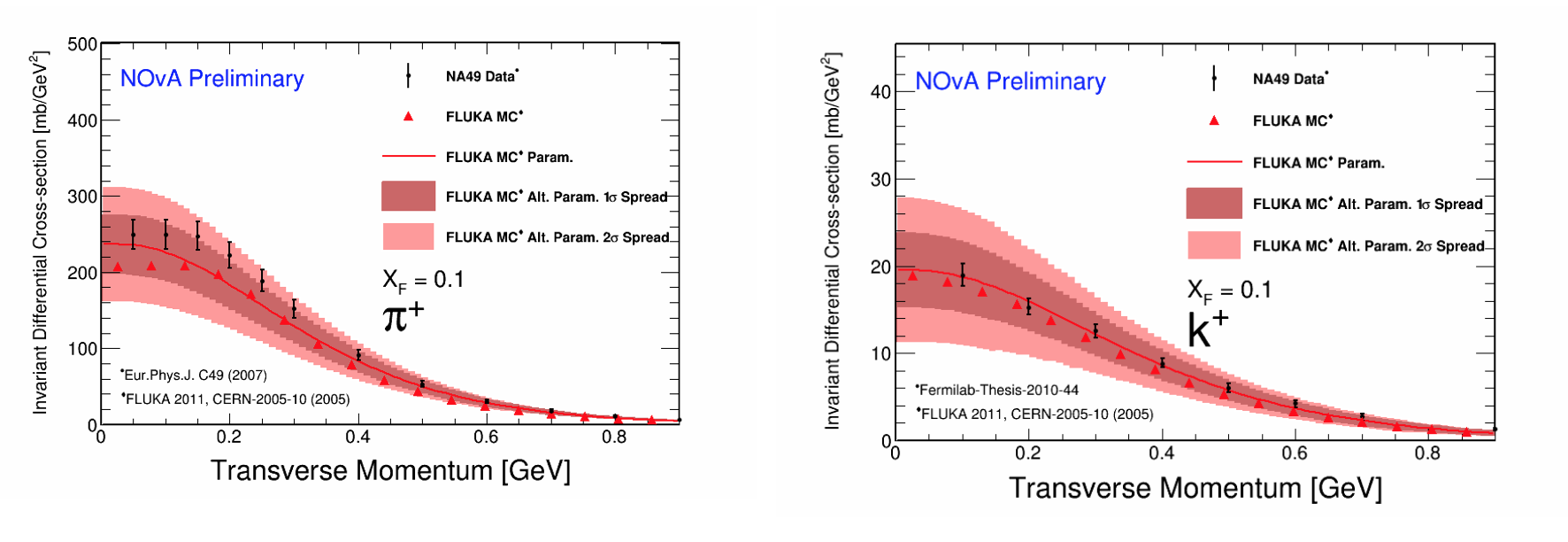}

 \caption{Invariant differential cross section for particular x$_{F}$ and as a function
of p$_{T}$ for Pions (left) and Kaons (right) produced in p+C collisions at 158 GeV/c beam momentum. Data is shown in solid black, MC in solid light red,
the parameterization of the MC as a light red dotted line, and the 1 $\&$ 2 sigma
spread in alternative parameterizations of the MC is shown as a pair of light
red bands.}\label{fig:p8}

 \end{figure}
 \begin{figure}[ht]
  \includegraphics[width=13cm]{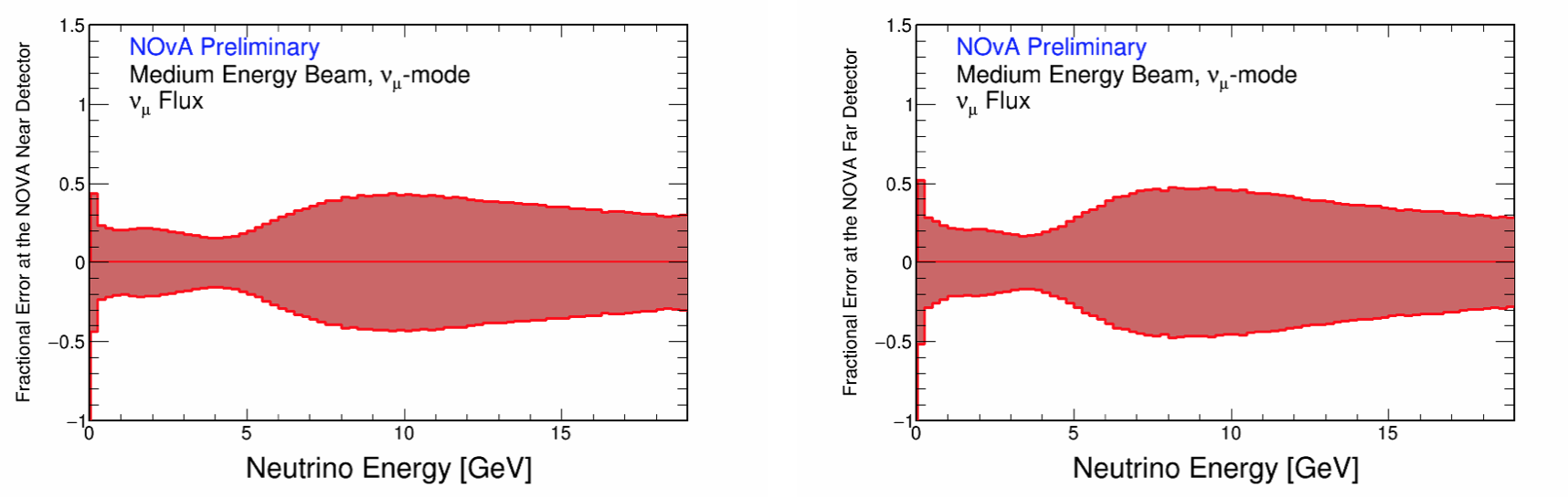}

 \caption{The square root of the diagonal elements of the covariance matrix describing the hadron production uncertainty on the beam $\nu_{\mu}$ flux at the ND and FD.}\label{fig:p9}
  \end{figure}

\begin{figure}[ht]
  \includegraphics[width=12cm]{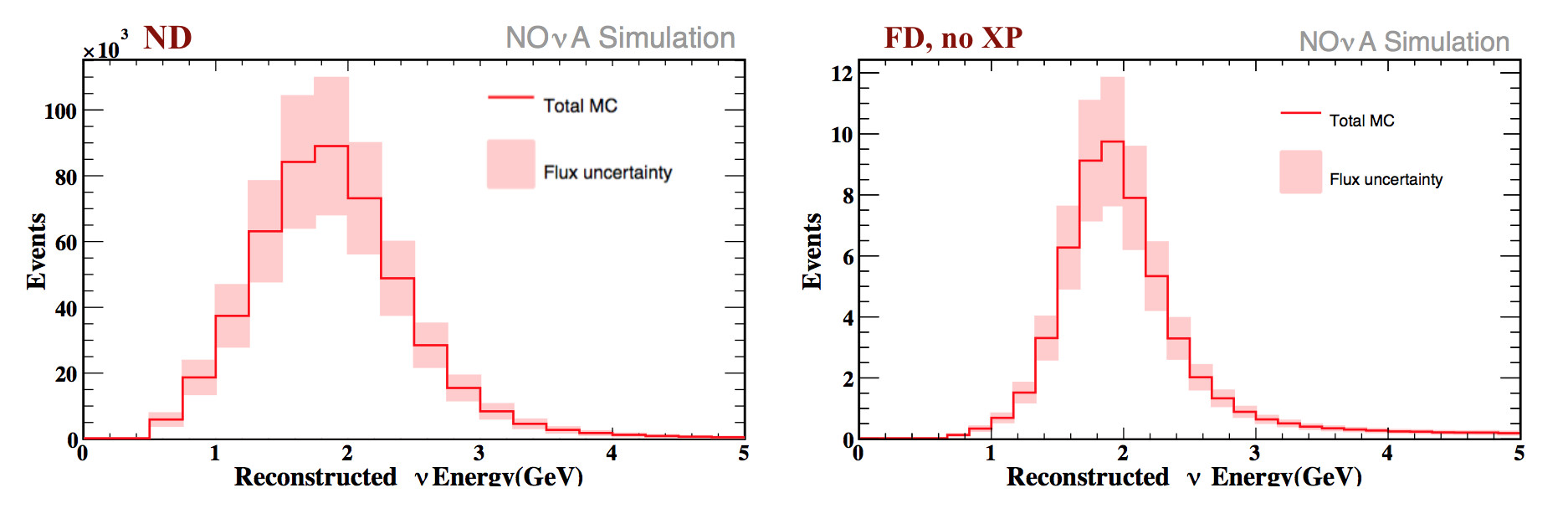}
  \caption{The error band represents a ±1 sigma shift of all beam systematics: including NA49 Hadroproduction Uncertainty, Spot size, Beam position on the target (X/Y), Target position, Horn current, Horn positions, $\&$ the modeling of horn$'$s B-field.}\label{fig:p10}

\end{figure}

The difference between nominal and shifted parameterizations is used to create
 weights in $P_T$ and $x_F$ of hadrons produced off the NuMI target, which, then, 
 can be used to re-weight the NO$\nu$A Near and Far Detector neutrino spectra, see 
 Figure~\ref{fig:p8} and Figure~\ref{fig:p9}.
 
Beam Transport Errors, including NA49 Hadroproduction Uncertainty on Reconstructed neutrino energy[GeV] in NO$\nu$A ND $\&$ FD for 6e20 POT as shown in Figure~\ref{fig:p10}.
\section{\bf Constraints using ND Data}
\begin{figure}[ht]

\begin{minipage}{.4\linewidth}
  \includegraphics[width=\linewidth]{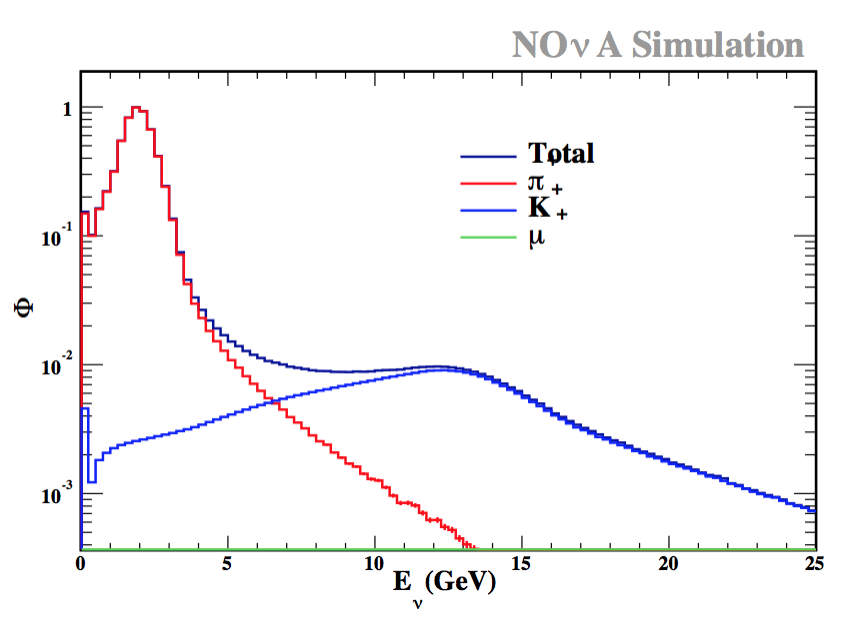}
  \caption{E$_{\nu_{\mu}}$ at ND}\label{fig:p11}
 
\end{minipage}
\hspace{.07\linewidth}
\begin{minipage}{.4\linewidth}
  \includegraphics[width=\linewidth]{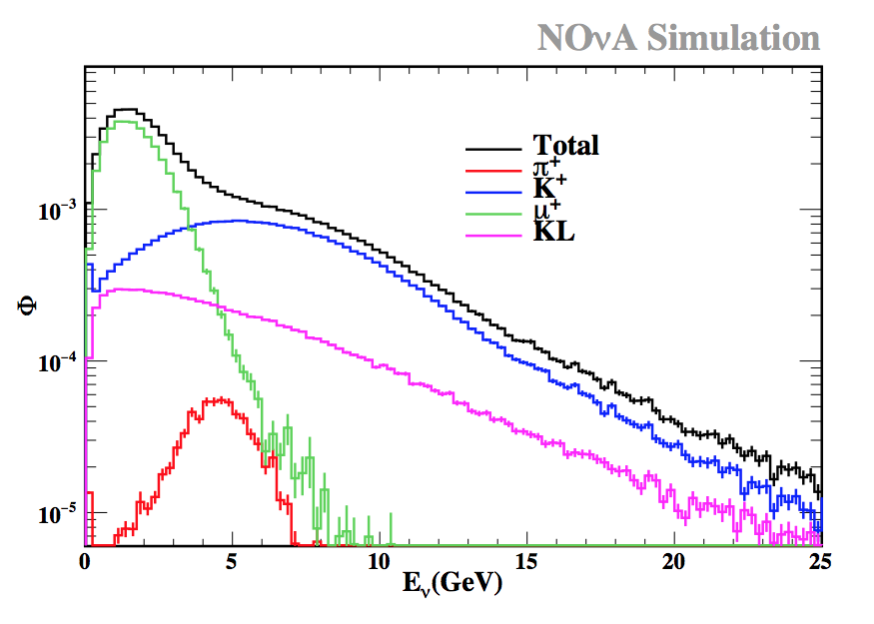}
  \caption{E$_{\nu_{e}}$ at ND}\label{fig:p12}
 
\end{minipage}
\end{figure}
We have shown systematic uncertainties from the beam transport and hadron production.
These predictions need to be further constrained by the ND Data. \\
Since 97$\%$ of $\nu_{\mu}$  at the ND are from 
$\pi\rightarrow\nu_{\mu} + \mu$.  We plan to use neutrino data in 1$\mbox{-}$3 GeV 
to constrain the pion yield, and use $E_\nu \geq 5$ GeV to constrain K$^{+}$ yield.

\section{\bf Summary:}
We present the flux systematic errors arising from the uncertainties in the beam transport  $\&$ hadro-production MC model. For beam transport parameters $\delta(\%)$ for $\nu_{\mu}, \bar{\nu_{\mu}}, \nu_{e}, \bar{\nu_{e}}$ is $\approx$3$\%$ for ND and FD(1$\mbox{-}$3)GeV, Energy variation for$\nu_{\mu}, \bar{\nu_{\mu}}, \nu_{e}, \bar{\nu_{e}}$ $\approx$1$\%$ for ND and FD(1$\mbox{-}$3GeV). Combined uncertainties using hadroproction with beam transport parameters at ND is $\pm$23.9$\%$ and  at ND is $\pm$20.9$\%$\cite{technote}. The flux prediction can be made more precise by using the
constraints provided by the neutrino spectra from ND, as we plan to do in the future.

\clearpage
\clearpage

\bibliographystyle{unsrt}
\bibliography{refs}

\end{document}